# Author Name Co-Mention Analysis:
# Testing a Poor Man's Author Co-Citation Analysis Method

Andreas Strotmann[1] and Arnim Bleier[2]

[1] *andreas.strotmann@gesis.org*
[2] *arnim.bleier@gesis.org*
GESIS – Leibniz Institute for the Social Sciences, Unter Sachsenhausen 6-8, D-50667 Cologne (Germany)

**Abstract**
As a social science information service for the German language countries, we document research projects, publications, and data in relevant fields. At the same time, we aim to provide well-founded bibliometric studies of these fields. Performing a citation analysis on an area of the German social sciences is, however, a serious challenge given the low and likely significantly biased coverage of these fields in the standard citation databases. Citations, and especially author citations, play a highly significant role in that literature, however.
In this work in progress, we report preliminary methods and results for an author name co-mention analysis of a large fragment of a particularly interesting corpus of German sociology: a quarter century's worth of the full-text proceedings of the Deutsche Gesellschaft für Soziologie (DGS), which celebrated its 100th anniversary meeting in 2012. Results are encouraging for this poor cousin of author co-citation analysis, but considerable refinements, especially of the underlying computational infrastructure for full-text analysis, appear advisable for full-scale deployment of this method.

**Conference Topic**
Visualisation and Science Mapping: Tools, Methods and Applications (Topic 8).

**Introduction**

In a way, this paper goes back to one of the very roots of author co-citation analysis, as White (1990) identified Rosengren's (1968) use of "co-mentions" of writers of literary works to visualize an intellectual structure in their reception.

Here, we take Rosengren's (1968) original writer co-mention visualization idea and extend it with the author citation pattern analysis and visualization methodology tradition of White and Griffith (1981), who extended the document co-citation idea generally co-attributed to Marshakova (1973) and Small (1973) to the study of author co-citation patterns, and of Zhao and Strotmann (2008a), who subsequently extended Kessler's (1963) bibliographic coupling idea from the study of document similarities to the study of author bibliographic coupling patterns with similar techniques.

From a technical perspective, the procedures used in this study are rooted in the idea of word co-occurrence analysis (a literature we will not review here), but with a restriction to words which likely denote frequently mentioned author names (or more precisely, their surnames). The analysis and visualization of author surname co-occurrence matrices (i.e., author name co-mention matrices) that we perform here uses methodology taken straight from Strotmann and Zhao (2012) – we refer the reader to that paper as a starting point for tracing its origins.

*Author Co-citation Analysis vs. Author Co-mention Analysis*

A formal author co-citation analysis of a document set usually proceeds as follows:
1. the reference lists of all documents are collected, usually from Web of Science;
2. for each cited reference found, names of authors of the cited work are identified (usually only the first author);
3. the authors cited most highly in the document set are determined;

4. an author × author co-citation matrix is constructed for the most highly cited authors in the document set – each cell counts the number of papers that are registered as co-citing the corresponding pair of authors;
5. the resulting co-citation matrix is analyzed statistically, and the result visualized and interpreted.

In the quarter-century of DGS proceedings that we selected for this experiment as a potential representation of "German sociology", unfortunately, formal *references* are frequently hard to identify, if they are listed at all. *Names* of sociologists (or other influential thinkers) are mentioned abundantly in these volumes, however: the surname of German sociologist Max Weber, to give a decidedly non-random example, appears at least once in every proceedings text on average. We therefore used *author co-mention analysis* as a poor-man's alternative to author co-citation analysis, as follows:

1. the full text of all documents is collected;
2. for each document full text, a list of candidate author surnames it mentions is compiled;
3. from the collection of all candidate author surnames mentioned, the most frequently mentioned likely surnames are extracted manually;
4. for each document, a weighted co-mention count is determined for each pair of frequently mentioned author surnames that it contains;
5. the per-document author surname co-mention matrices are accumulated into a corpus-level surname × surname co-mention matrix;
6. the co-mention matrix is analysed statistically, and the results are visualized and interpreted.

*The DGS 1960-85 corpus*

The corpus we use in this experiment is a large fragment (about 25%) of a particularly interesting series of publications in German sociology: a quarter century's worth of the full-text proceedings of the roughly biannual meetings of the *Deutsche Gesellschaft für Soziologie* (DGS), which celebrated its 100[th] anniversary meeting in 2012. Our experiment is performed in part to determine if it is possible to perform meaningful citation analysis on this corpus, and to determine areas in which new methods may need to be developed.

The DGS proceedings used for the experiment had previously been scanned, OCRed and catalogued by GESIS Leibniz Institute for the Social Sciences; where possible, this corpus has been made available to the general public on the institution's Social Sciences Open Access Repository (SSOAR.info). The particular fragment of this corpus that we used in this experiment is comprised of 1,212 publications which appeared during the years 1960 to 1985. On average, each document contains 2,471 words, mostly in German.

*Identifying frequently mentioned author surnames*

In many languages, names are the only words that are capitalized inside a sentence, so that the problem of extracting names from a text is largely reduced to filtering out corporate names. In German orthography, however, which almost all the texts in our corpus adhere to, all nouns are capitalized, not just names, and surnames especially are taken from a wide range of concept nouns (e.g., Vogel = bird; Weber = weaver), place names (e.g., Mannheim), or Christian names (e.g., Walter), each from a range of languages, regions, and cultures. This is further complicated by two forms of attribution suffixes in German, namely, the genitive case marker ("Freuds" for English "Freud's") and its "sch" suffix form ("Freudsche/n/m/r" for English "Freudian"). In our experiment we decided for simplicity's sake to use words as text units rather than compound phrases, which exacerbates the name ambiguity problem.

Identification of author name mentions is thus a major problem in our case, which we addressed only approximately for the sake of this experiment. The approach we opted for was to create a list of words that count as mentions of authors, and to treat everything that does not appear in this list as non-name words. As central criteria for the construction of this list, we would like central and frequently mentioned authors to be included, but at least in the current experiment would prefer to remove from the list any terms that frequently occur as concept words rather than names. We would also like to remove terms that are too likely to name many different individual authors.

We chose to approximate these criteria by creating the list of likely author surnames in the following steps: First, we compiled a list of candidate author surnames from the author metadata of a document collection large enough that it can be assumed to have a similar distribution of surname frequencies as the target corpus. Next, we pruned from the resulting list those names that do not begin with a capital letter, that are very short, or that occur only rarely. We lemmatized this list of names (removing genitive markers and the like as discussed above). Finally, candidate names too likely to refer to multiple authors or to non-author entities were removed.

For the purpose of this experiment the SOWIPORT document collection (see, e.g., Stempfhuber, 2008) presented itself as an attractive source from which to compile the list of candidate author surnames. SOWIPORT covers the German social sciences, something of a superset of our target distribution of author name mentions in the DGS corpus, and it makes authorship metadata readily available. For the subsequent pruning step, we found experimentally that limiting the list of candidate author names to words of at least three characters which occur at least 25 times in the DGS corpus yielded useful results in our setting. We applied lemmatization following the flexion rules of German proper nouns. This resulted in a list of about 1500 different words that frequently appear as author surnames in SOWIPORT and frequently appear in our target corpus. From this list we first hand-picked about 500 to proceed with, lemmatized as above. Finally, we manually removed about another half as being immediately recognizable as concept words, leaving about 220 names in total, each mentioned at least 11 times in the corpus.

*Constructing the author name co-mention matrix*

Next, for each member in our list the full texts were scanned for lemmatized word matches, resulting in 220 multisets of document author mentions, i.e., one multiset of documents per lemmatized author surname, with each document occurring in the multiset as many times as the corresponding name appears in the document.

Clearly, this list of words, which hopefully denote unique authors cited frequently in German sociology during the years 1960-85, does not qualify as a complete list of authors, nor is it a random sample, since highly cited names are selected preferentially. However, experience teaches us that author co-citation data visualized through factor analysis is quite a stable technique, which means that there is a good chance that intelligible results could be obtained from applying a similar method to our target corpus even with a sub-optimal list of names.

Unlike traditional co-citation analysis based on data from citation databases, where each cited publication appears once in the list of references even if the text refers to it dozens of times, co-mentions between authors allow for a weighting by how many times each name appears in the text. Intuitively, a document that refers frequently to two authors indicates a stronger connection between them than a document that either mentions both authors just once or that mentions one author dozens of times while the other author is mentioned just once.

To calculate the co-mention count of two authors in this set of documents, we therefore first calculate for each author his or her author mention profile as the multiset (rather than the set) of documents that mention the author, weighted by the number of times that author is mentioned in the document. Given these author mention profiles for all the frequently mentioned author surnames identified in the previous step, the co-mention count of two authors is calculated as the multiset size of the multiset intersection of the two authors' author mention profiles.

The typical author co-citation count for a pair of authors would be determined as the size of the intersection of the sets (not multisets) of documents that cite each author. The inspiration for our choice of co-mention counting method is Zhao and Strotmann (2008a), who calculate the author bibliographic coupling of two authors as the multiset size of the multiset intersection of the citing behaviour profiles of the two authors, where an author's citing behaviour profile is the multiset union of all reference lists of the author's oeuvre.

Unlike in the case of all-author co-citation matrices (Zhao, 2006; Zhao & Strotmann, 2008b), it is not possible to distinguish between inclusive and exclusive co-mention counts, i.e., to filter out co-mentions based purely on co-authorship (e.g., Marx & Engels). Indeed, co-mention counts in this experiment even include the co-occurrence between two authors of the mentioning work, and co-occurrences of the names of the mentioning work's authors and the mentioned work's authors. Author co-mention matrices are thus considerably more noisy data sources than author co-citation matrices.

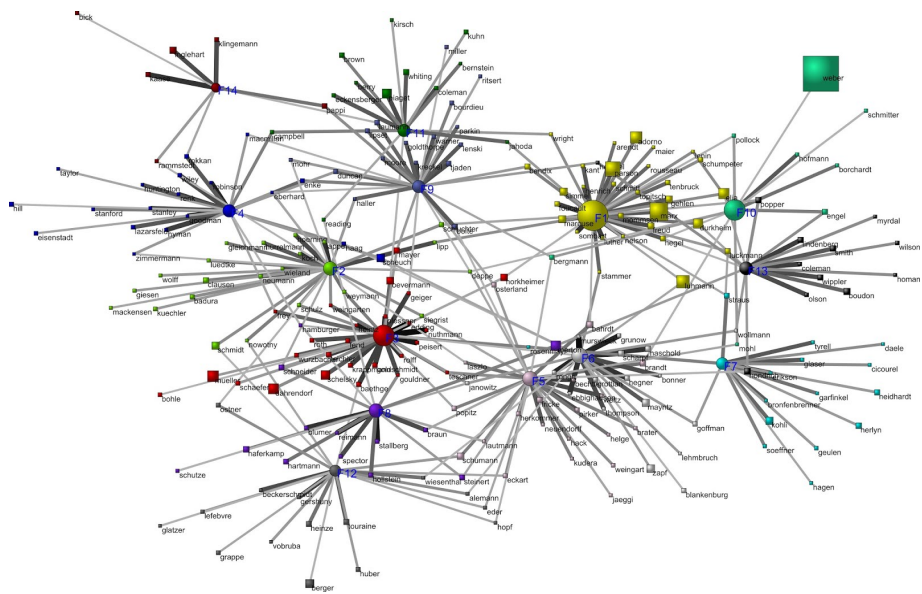

Figure 2: Author name co-mention analysis result visualization, 14 factor solution.

*Factor analysis and visualization*

Finally, we performed an exploratory factor analysis on the resulting author co-mention matrix, using the Factor Analysis routine of SPSS 19. As parameters, we specified: replacement of missing values (i.e., diagonal values) by the mean; oblique rotation using Oblimin with default parameters; and a maximum of 250 iterations for each of the steps involved in the factor analysis.

Based on visual inspection of the Scree plot, we chose a 14-factor solution rather than the 38-factor one that Kaiser's Rule of eigenvalue > 1 would have suggested. We visualize the 14 factors × 220 authors pattern matrix result of the factor analysis in Figure 2. In this visualization using the Kamada-Kawai layouting routine of Pajek 3, lemmatized author names and factors are represented as nodes (square or round, resp.), and an author's loading on a factor is represented by a line with gray scale value and width proportional to the absolute loading. Nodes are color coded according to the 14 factors and their memberships (authors who do not load sufficiently on any factor are not displayed). Author node sizes are proportional to the number of mentions, and factor node sizes proportional to the sum of members' mentions weighted by the member's factor loading.

*Interpretation*

The factors were interpreted and labelled by two social scientists, both colleagues of the authors. Table 1 lists the results of this interpretation, along with some statistical characteristics of the factors: the size of a factor (defined as the number of authors whose maximal loading (of at least 0.3) in the pattern matrix is with this factor) and the highest loading of an author on this factor (which is an indicator of the clarity or distinctness of this factor).

Factor labels ending in a question mark denote factors for which an intellectual interpretation was not easily apparent, and for which an hour's discussion between the social scientists did not lead to a clear agreement. A label for these factors was attempted by the authors using SOWIPORT and Google Scholar in these cases. Generally, though not always, these more "questionable" factors tend to exhibit lower maximal author loadings, as Table 1 shows, and the most questionable one, F10, identified as an artefact of the lack of author name disambiguation in the underlying dataset, has the lowest such characteristic.

Table 1. Factors and their interpretations and characteristics.

| *Factor* | *Label* | *Max Loading* | *# Members* |
|---|---|---|---|
| F1 | Theory of Society | .87 | 30 |
| F2 | Biographies? | .79 | 22 |
| F3 | Government Theory | .97 | 26 |
| F4 | Political Science | .85 | 19 |
| F5 | Sociology of Work? | .75 | 20 |
| F6 | Sociology of Organisations | .92 | 17 |
| F7 | Sociology of the Family? | .77 | 15 |
| F8 | Social Problems? | .90 | 13 |
| F9 | Social Inequality? | .70 | 16 |
| F10 | ?common names? | .63 | 8 |
| F11 | Psychology? | .72 | 12 |
| F12 | Socioeconomics? | .76 | 15 |
| F13 | Rational Choice Theory | .76 | 12 |
| F14 | Values and elections? | .80 | 6 |

**Discussion and Outlook**

In this paper we present a first experiment using author name co-mention analysis based on the full text of a scientific literature for which no formal citation index and therefore no possibility of a traditional author co-citation analysis is available.

Given the quality of the underlying dataset we used here, the results are encouraging. Statistical characteristics of the factor analysis results are reasonable when compared to those from author co-citation analyses of other fields we have performed previously. Interpretation of about half the resulting factors was considered straightforward by the social scientists who looked at the results; again, this suggests reasonable performance in our experience.

The appearance of a factor (F10) that consists mostly of surnames that likely correspond to several distinct authors each reminds us, however, that the author name co-mention methodology we tried here has its limits. Author name disambiguation is a requirement that we attempted to avoid by filtering out words that could, in principle, be either surnames, first names, or dictionary words, or any combination of these. As Strotmann and Zhao (2012) point out in a similar case, this is not always a reasonable approach to take to author name disambiguation, and for author name co-mention analysis to work well, significant effort will need to be invested in improving the identification of individuals from author name mentions. The fact, on the other hand, that the name "Weber" - almost certainly denoting, in a vast majority of cases, the prominent founding father of sociology in Germany, Max Weber - gets categorized with these multi-individual names is perhaps symptomatic of the extreme degree to which the founders of German sociology are cited across its entire literature. The size of the Weber author node in our visualization - at least an order of magnitude larger than even the factor nodes - illustrates this, too: the name *Weber* is practically synonymous with the term "sociology", being mentioned more than 2000 times in 1200 texts.

We suspect that one reason why this experiment worked quite well despite these shortcomings is the fact that author name co-mention counting allowed us to weight higher those who are (co-)mentioned frequently in a text as opposed to those who are mentioned only in passing. This would be expected to significantly increase the relevance of high co-mention counts, and thus improve the signal-noise ratio.

Especially for the purposes of bibliometric analysis of social science literatures, where coverage of standard citation indexes are considered inadequate, this alternative approach may serve as a poor-man's analysis tool as long as the data situation does not improve significantly.

## Acknowledgments

The authors wish to thank Howard D. White and Dangzhi Zhao for excellent advice on this paper, and their GESIS colleagues Maria Zens and Matthias Stahl for help interpreting the factor analysis results.

## References

Kessler, M.M. (1963). Bibliographic coupling between scientific papers. *American Documentation, 14*, 10-25.

Marshakova, I.V.(1973). System of document connections based on references (in Russian). *Nauchno-Tekhnicheskaya Informatsiya, 2*(6), 3-8.

Rosengren, K.E. (1968). *Sociological aspects of the literary system*. Stockholm: Natur och Kultur.

Small, H. (1973). Co-citation in the scientific literature: A new measure of the relationship between two documents. *Journal of the American Society for Information Science, 24*, 265.

Stempfhuber, M., Schaer, P., & Shen, W. (2008). Enhancing visibility: Integrating grey literature in the SOWIPORT Information Cycle. *The Grey Journal, 4*(3), 121.

Strotmann, A., & Zhao, D. (2012). Author name disambiguation: What difference does it make in author-based citation analysis? *Journal of the American Society for Information Science, 24*, 265.

White, H.D. (1990). Author co-citation analysis: overview and defense. In *Bibliometrics and Scholarly Communication*, Christine Borgman, ed. Newbury Park, CA: Sage. 84-106.

White, H.D., & Griffith, B.C. (1981) Author co-citation: a literature measure of intellectual structure, *Journal of the American Society for Information Science, 32*, 163-172.

Zhao, D. (2006). Towards all-author co-citation analysis. *Information Processing & Management, 42*, 1578.

Zhao, D., & Strotmann, A. (2008a). Evolution of research activities and intellectual influences in Information Science 1996-2005: Introducing author bibliographic coupling analysis. *Journal of the American Society for Information Science and Technology, 59*(13), 2070.

Zhao, D., & Strotmann, A. (2008b). Comparing all-author and first-author co-citation analyses of Information Science. *Journal of Informetrics, 2*(3), 229.